\documentclass[12pt]{iopart}
\usepackage{graphicx}
\usepackage{setspace}

\graphicspath{{Figures/},.}
\begin{document}

%----------------------------------------------------------------------
%  TITLE and AUTHORS
%----------------------------------------------------------------------
%\title[Optical properties and characterisation of nanostructured
%GaSb]{Optical properties and characterisation of nanostructured GaSb}

\title[Characterisation of nanostructured GaSb]{Characterisation of
  nanostructured GaSb : Comparison between large-area optical and
  local direct microscopic techniques}

\author{I.S.~Nerb\o$^{1*}$, M.~Kildemo$^1$,
  S.~Leroy$^2$, I.~Simonsen$^{1,2}$, E.~S\o ønderg\aa rd$^2$,
  L.~Holt$^1$ and J.C.~Walmsley$^1$}

\address{$^1$Department of Physics, Norwegian University of Science
  and Technology~(NTNU),
  NO-7491 Trondheim, Norway\\
  $^2$UMR 125 Unité mixte CNRS/Saint-Gobain Laboratoire Surface du
  Verre et Interfaces 39 Quai Lucien Lefranc, F-93303 Aubervilliers
  Cedex, France } \ead{ingar.nerbo@ntnu.no}

% --- The abstract ---
\begin{abstract}
Low energy ion-beam sputtering of GaSb results in
self-organized nanostructures, with the potential of structuring large
surface areas. Characterisation of such nanostructures by optical methods
is studied and compared to direct (local) microscopic
methods. The samples consist of densely packed GaSb cones on bulk
GaSb, approximately $30$, $50$ and $300 \mbox{nm}$ in height, prepared by sputtering
at normal incidence. The optical properties are studied by
spectroscopic ellipsometry, in the range $0.6$--$6.5 \mbox{eV}$, and with Mueller
matrix ellipsometry in the visible range, $1.46$--$2.88 \mbox{eV}$. The optical
measurements are compared to direct topography measurements obtained
by Scanning Electron Microscopy (SEM), High Resolution Transmission
Electron Microscopy (HR-TEM), and Atomic Force Microscopy (AFM). Good
agreement is achieved between the two classes of methods when the
experimental optical response of the short cones ($<55 \mbox{nm}$) is inverted
with respect to topological surface information via a graded
anisotropic effective medium model. The main topological parameter measured was  the average cone height, but estimates of typical cone shape and density (in terms of volume
fractions) were also obtained. The graded anisotropic effective medium
model treats the cones as a stack of concentric cylinders (discs) of
non-increasing radii. The longer cones ( $300 \mbox{nm}$) were found
experimentally to give rise to considerable reduction in the measured
degree of polarisation. This is presumably caused by multiple
scattering effects, and thereby questions in this case the validity of
the effective medium approximation. Moreover, the reflectance of the
samples were also measured and modelled, and found to be reduced
significantly due to the nano-structuration. Optical methods are shown
to represent a valuable characterization tool of nanostructured
surfaces, in particular when large coverage area is desirable. Due to
the fast and non-destructive properties of optical techniques, they may
readily be adapted to {\em in-situ} configurations.
\end{abstract}

% Uncomment for PACS numbers title message \pacs{00.00, 20.00, 42.10}
% Keywords required only for MST, PB, PMB, PM, JOA, JOB?  \vspace{2pc}
\maketitle

%--------------------------------------------------------------------
%   MAIN TEXT
%--------------------------------------------------------------------

% --- Introduction
% -------------------------
\section{Introduction}

 The recent advent of nano-technology and nano-science has made
  it increasingly important to be able ``see'' features of a sample
  down to a nanometric scale. This is today typically achieved with
  the aid of several well-established microscopic
  techniques like, Atomic Force Microscopy~(AFM), Scanning
  Electron Microscopy~(SEM) and Transmission Electron
  Microscopy~(TEM). All these techniques can achieve
  nanometric resolution, and are therefore attractive when one wants to
  image the fine details of a sample. Moreover, they can
  be said to be {\em local} in the sense that only a rather small
  surface area can be imaged with good resolution. They are also
  rather time consuming techniques, and the required equipment is
  costly and physically large. As a result, they are not generally suited for {\em in-situ}
  characterisation.

Traditionally, optical techniques have been attractive for {\em
    in-situ} studies due to its measurement speed, relatively low
  equipment cost, non-contact properties and ease of integration with
  other setups. Optical techniques also have the advantage of being
  able to cover large surface area with relative ease. This is a great
  advantage, for instance, in monitoring applications where it is the
  average properties that are of interest, and not the local features
  at a given location at the surface. For nanometre scale
  structures, the applicability of optical techniques are limited by
  the diffraction limit~\cite{BornWolf}, making imaging of such
  structures by visible light impossible.  However, even if direct
  optical imaging is challenging for nano-structures, it is well-known
  that they can have strong polarisation altering properties on the
  incident radiation. Hence, indirect optical techniques can in
  principle be devised for the purpose of extracting topographic
  information about the sample. The aim of this paper is to present
  such a methodology, and to compare the large area optical
  result to local information obtained by direct methods.

Spectroscopic ellipsometry~(SE) is a celebrated polarimetric
technique, much used for measuring the thickness of thin film layers,
and for determining the index of refraction of materials. It can also
be used to characterise nanostructures. For example, generalised ellipsometry has been used to measure the inclination angle of
nano-rods~\cite{Robbie}. 
The sensitivity of spectroscopic ellipsometry to the thickness of thin layers is remarkable, and can be down to
single atom layers. This is achieved by knowing the refractive indices
of the materials, and utilising optical models. The aim of this study
is to exploit SE's sensitivity of thin film thickness, in order to accurately
measure the height of conical shaped nanostructures. This is 
done by developing a suitable optical model. Information on shape and
regularity can also possibly be attained. The ellipsometric spot will
always average over a relatively large (surface) area, providing
information on the mean properties of the structures. It is both
non-invasive and fast, making it suitable as a tool for
\emph{in-situ} characterisation.

Nanostructured surfaces and materials open up for a new range of
applications. In photonics, for example, optical
properties of thin films may be mimicked by
nanostructures, and further supply new and enhanced properties (see
{\it e.g.}~\cite{Robbie}, and references therein). An example of such
properties can be anti-reflective coatings
with low reflectivity over a large spectral range
and a wide range of incident angles~\cite{Zu-Po}.

Low energy ion sputtered GaSb is a good example
of self-organized formation of densely packed cones, and has been
proposed as a cost-effective method for production of {\it e.g.} quantum
dots~\cite{Facsko}.  The properties of such a surface may, to a large
extent, be tailored by controlling sputtering conditions. The latter
issue is a typical target application area of ellipsometry. In the case of
a future large scale production of such surfaces, SE could possibly be
used as an efficient production control tool for testing individual
samples.

Here, optical models are initially constructed from observations from
High Resolution Transmission Electron Microscopy~(HR-TEM), Field
Emission Gun Scanning Electron Microscopy~(FEG-SEM), and AFM. The
latter give a direct observation of the nanostructures, with respect
to density, cone separation, cone height, number of nearest neighbours
etc. Information on the shape and crystal structure of individual cones, were obtained from HR-TEM studies of carefully prepared slices of
nanostructured GaSb.

This paper is organized as follows: In section 2 we describe the experimental details of both the direct microscopic (SEM, TEM and AFM), and the optical (SE) studies. A brief theoretical background on ellipsometry is also given. In section 3 we present the results of these studies. The optical properties of conical nanostructures are discussed in relation to the effective medium approximation. An optical model is presented in section 3.3, that enables characterisation of such structures from optical measurements, by fitting the model parameters to the SE measurements.Information gained from the optical characterisation are finally compared to the results from the direct microscopic studies. 

%To be able to characterise nanostructures from optical measurements, an optical model is needed. We present such a model in section 3.1, and fit its parameters to the SE measurements. 

\section{Experimental details and theoretical background}

The samples consisted of GaSb sputtered at
low-ion-energy~\cite{Sondergard}. The deposition condition for each
sample is reported in table~\ref{tab:samples}. The substrates were
crystalline GaSb~($100$) wafers, of thickness $ 500 \mbox{$\mu$m}$. All
samples were sputtered at room temperature.  The FEG-SEM images were
obtained using a Hitachi S-4300SE, and Zeiss Supra instruments. TEM
analysis was performed using a JEOL 2010F. Cross section TEM samples
were prepared by both ion-milling and ultramicrotomy, in order to
investigate possible preparation induced artefacts in the
microstructure. Ion-milling was performed using a Gatan PIPS
instrument, operating at $3.5 \mbox{kV}$ with a thinning angle of
$3.5$--$4^\circ$. Ultramicrotomy was performed using a Reichart-Jung
Ultracut E instrument. AFM measurements were done by a DI-VEECO AFM with a NanoScope IIIa controller from Digital Instruments, operated in tapping mode. Silicon tips with radius less than $10 \mbox{nm}$ were used. 

The optical far field measurements were performed using a commercial
Photo-Elastic-Modulator Spectroscopic Ellipsometer~(PMSE) in the
photon energy range $0.6$--$6.5 \mbox{eV}$ (UVISEL, Horriba Jobin
Yvon), at an angle of incidence of $55^\circ$. The complete Mueller
matrix was also measured using a commercial ferroelectric
liquid-crystal retarder-based Mueller matrix Ellipsometer~(MM16) in
the range $1.46$--$2.88 \mbox{eV}$ ($850$--$430 \mbox{nm}$). Such measurements where made for several angles of incidence in the
range from $55^\circ$ to $70^\circ$, and as a function of the
sample rotation angle around its normal to the (mean) surface. The
orientation of the sample with respect to the direction of the
incoming beam was carefully recorded, and the sample was rotated
manually in steps of $45(\pm2)^\circ$, with a total sample rotation in
all cases of at least $360^\circ$.

The PMSE measurements were performed in the standard UVISEL set-up,
{\it i.e.} polariser-sample-PEM-analyser, where the angle of the fast
axis of the PEM with respect to the analyser is fixed at $45^\circ$.
Measurements were performed in the standard PMSE configurations
($\Theta_M=0^\circ$, $\Theta_A=45^\circ$)~\cite{Drévillon}, determining
$I_s=-m_{43}$ and $I_{c}=m_{33}$, where $m_{43}$ and 
$m_{33}$ are normalised Mueller matrix elements related to the
unormalised Mueller matrix $M$ by $m=M/M_{11}$. 
For the reflection from a isotropic planar surface, they can be
defined according to
\begin{eqnarray}
  I_s= -m_{43} = \sin2\Psi \sin\Delta, 
  \label{eq:Is}
\end{eqnarray}
and 
\begin{eqnarray}
  I_c= m_{33} = \sin2\Psi \cos\Delta,
  \label{eq:Ic}
\end{eqnarray}
where $\Psi$ and $\Delta$ denote the ellispometric angles related to the ratio of the complex reflection amplitudes
$r_{pp}/r_{ss} = \tan\Psi e^{i\Delta}$~\cite{Hauge}.
%\begin{eqnarray}
%  \frac{r_{pp}}{r_{ss}} = \tan\Psi e^{i\Delta}.
%\end{eqnarray}

Additional measurements were performed in the configuration
($\Theta_M=45^\circ$, $\Theta_A=45^\circ$), enabling the determination of
\begin{eqnarray}
  I_{c_{2}}=-m_{12}=\cos2\Psi.
  \label{eq:Ic2}
\end{eqnarray}
The quantities, $I_s$, $I_c$ and $I_{c_{2}}$, as defined in
equations~(\ref{eq:Is})--(\ref{eq:Ic2}) are known as the
  ellipsometric intensities. For block-diagonal Mueller matrices, these intensities can be used to define the {\em degree of polarisation}, $P$.~\cite{Jellison-McCamy}
  
\begin{equation}
  P=\sqrt{I_s^2+I_{c}^2+I_{c_{2}}^2}.
  \label{DOP}
\end{equation}
A discussion on when a sample will have a block-diagonal Mueller matrix will be given in the following section. From the full Mueller matrix, experimentally available here in the
range $1.46$--$2.88 \mbox{eV}$, it is also customary to define the
so-called {\em depolarisation index}~\footnote{Consult {\it e.g.}
  Ref.~\cite{Chipman:05} for a more detailed discussion of
  depolarisation measures.}
\begin{equation}
  D_P=\frac{\sum_{i,j} M_{ij}^2-M_{11}^2}{\sqrt3 M_{11}},
  \label{eq:DP}
\end{equation}
where $M_{ij}$ denotes the non-normalised Mueller matrix
elements~\cite{Chipman:05}.  
$P$ (and in most cases $D_P$) determine how much of the outgoing light will be totally polarised for totally polarised incident light.
"`Reflectance measurements"' were also
performed by the PMSE, in which $M_{11}=(R_{ss}+R_{pp}+R_{sp}+R_{ps})/2$ was
determined by using a standard Al mirror reference sample, and
assuming stable intensity conditions. The reflectance spectrum was
recorded from $1.5$--$6.5 \mbox{eV}$ in steps of $0.1 \mbox{eV}$.

\section{Results and discussion}
\subsection{Experimental observations --- SEM, TEM and AFM results}

\begin{figure}
  \centering
  \includegraphics[width=0.5\textwidth]{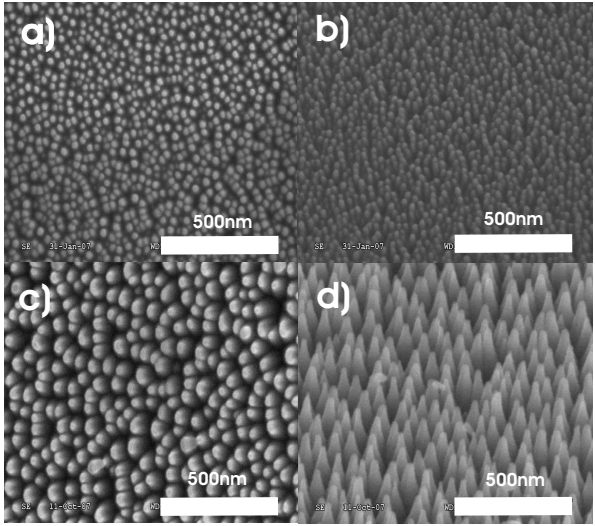}
  \caption{SEM images of GaSb nanocones. Figure (a), sample $A$ at
    normal beam incidence, (b) tilted sample $A$. Sample $D$ is also depicted at (c) normal beam 	incidence 
    and (d) tilted beam incidence }
  \label{fig:SEM}
\end{figure}

In figure~\ref{fig:SEM}, FEG-SEM images of samples $A$
and $D$ are presented, for both normal beam incidence (left-hand
images), and tilted beam incidence (right-hand images). The cones do not show a high degree of organisation, but on average have 6 nearest neighbours. This result was found from statistical treatment of AFM measurements of the samples, but could as well have been found from SEM images. The result
correspond well with Euler's law~\cite{Weaire}, which state that the
mean number of nearest neighbours for a structure created by a random
process is $6$. The average cone separation, $\langle D\rangle$, has
been estimated from the cone density, by assuming the cones to be
ordered on a perfect hexagonal lattice. The average height of the
cones, $\langle h \rangle$, were nominally estimated from AFM, but for
sample $A$, 
it was estimated from HR-TEM. The average cone heights and cone
separations are given in table~\ref{tab:AFM}, along with the estimated
standard deviation $\sigma_h$ of $\langle h \rangle$.

\begin{figure}
  \centering
  \includegraphics[width=\textwidth]{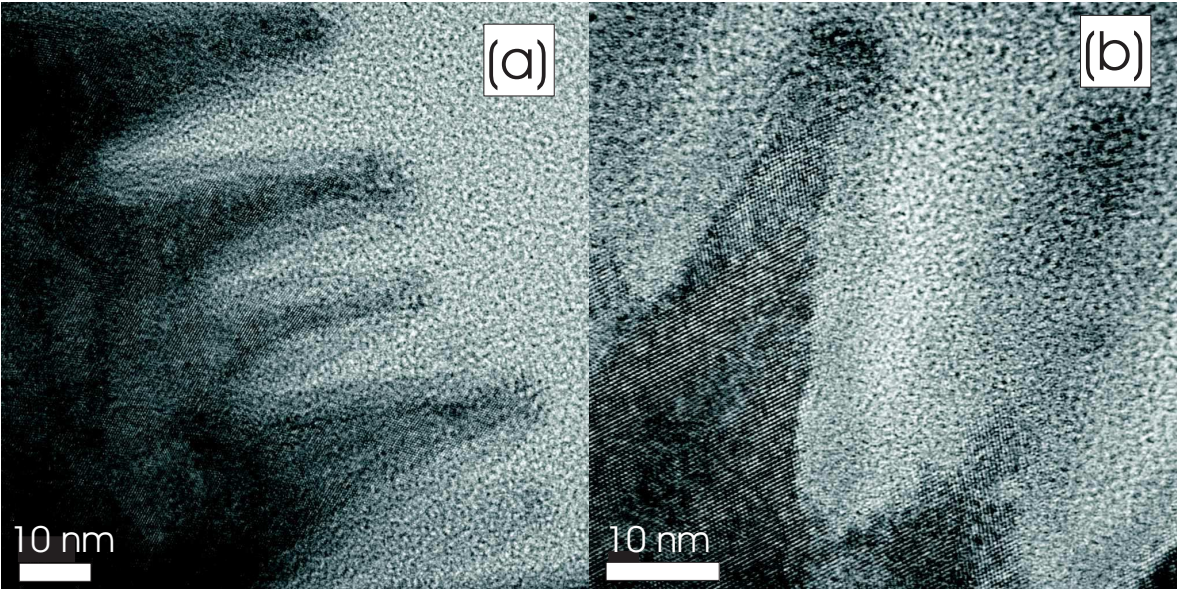}
  \caption{High resolution TEM images of GaSb nanocones (sample A). Figure (a) shows several cones while figure (b) shows one individual cone in greater detail. The lighter part of the image is the amorphous adhesive used in the sample preparation. The crystalline cones appear darker and in figure \ref{fig:TEM_IM}(b) the atomic column spacing at the 110 GaSb zone-axis orientation is clearly visible.  The approximately $5 \mbox{nm}$ layer of amorphous GaSb/oxide is visible as a shadow around the cones that has slightly darker contrast than the adhesive.}
  \label{fig:TEM_IM}
\end{figure}

Figure~\ref{fig:TEM_IM} depicts a HR-TEM image of selected cones from
sample A, prepared by ion-milling. The average
cone height of A was estimated to be $55
\mbox{nm}$, obtained by taking the average of $16$ cones measured by
HR-TEM. The shape was found to be conical with a somewhat rounded tip.
The typical cone angle, defined as the angle between the substrate and
the cone side wall, was found to be roughly $\alpha=73^\circ$.  From the
HR-TEM images in figure~\ref{fig:TEM_IM}, it is observed that the
interior of the cones consisted of primarily crystalline material,
with the same crystal orientation as the substrate. Furthermore, the
cone surface appeared to be surrounded by a thin (less than $5
\mbox{nm}$) layer of undetermined amorphous material.
From the rapid
oxidation of clean GaSb to an approximately $5$--$7 \mbox{nm}$
GaSb-oxide layer, it is argued that this surrounding layer is partially oxidised.

Another slice of nanostructured GaSb (sample $A$)
, was prepared by ultramicrotomy. This sample did
not provide such a thin sector as the
ion-milled samples.  However, it was sufficient to confirm the structure observed in the ion-milled samples. The crystalline
nature of the interior of the cones, and the amorphous surrounding
layer is in line with the observations by Fascko {\it et
 al.}~\cite{Facsko}.
%The crystalline nature of the cones, was less
%well observed from the ultramicrotomy (more diffuse images) than from the ion-milled images,
%see figure~\ref{fig:TEM_M}, although the trend appears to be similar.

In summary, from the TEM studies mainly three phases
appear to be involved in the layer (thin film) defined by the cones.
These phases are crystalline GaSb~(c-GaSb), amorphous GaSb~(a-GaSb) and
presumed GaSb-oxide, in addition to the voids between the cones.
Theremaining samples were studied by AFM and by FEG-SEM, and detailed results are compiled in table~\ref{tab:AFM}. 
It is suspected that the AFM tip is too blunt to reach the bottom between close-packed cones. Therefore, when estimating the average cone height $\langle h \rangle$, the height of each cone top have been defined relative to the lowest point in an area within the maximal distance between the cones. This minimum is typically found in a place where the cones stand further apart, and the tip can reach the bottom. This may overestimate the average height $\langle h \rangle$ somewhat.

\subsection{Experimental observations --- Spectroscopic Ellipsometry}

\begin{figure}
  \centering
  \includegraphics[width=0.5\textwidth]{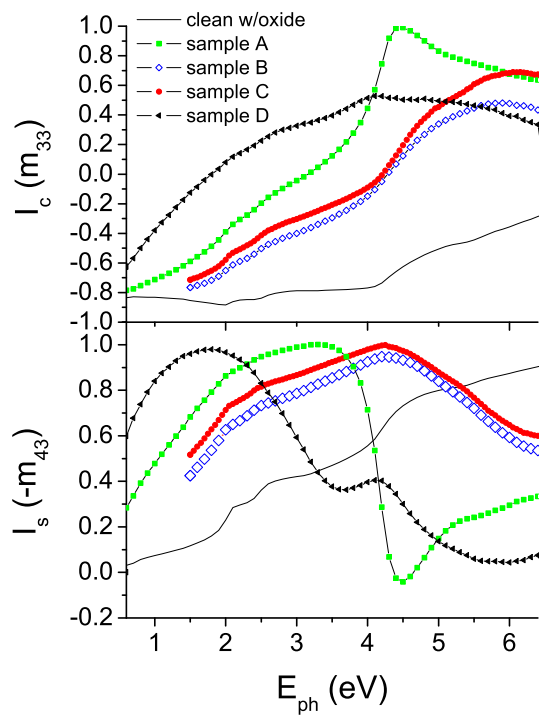}
  \caption{Ellipsometric intensities $I_s (-m_{43})$, $I_{c}(m_{33})$,
    of plane GaSb with 7 nm oxide, short nanostructured cones, samples
    $A$, $B$ and $C$ (around $50 \mbox{nm}$ high cones), and sample
    $D$ (approx. $300 \mbox{nm}$ cones).}
  \label{fig:IsIc}
\end{figure}

\begin{figure}
  \centering
  \includegraphics[width=0.5\textwidth]{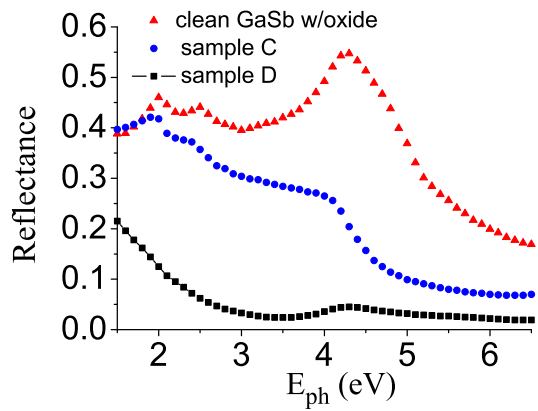}
  \caption{Reflectance ($M_{11}$) of nanostructured GaSb cones, for
    sample $C$ (approx 36 nm high cones) and $D$ (approx 300 nm high
    cones). As a reference, the reflectance of a clean GaSb surface
    with oxide is also included}
  \label{fig:R}
\end{figure}

Figure~\ref{fig:IsIc} shows the SE measurements of
$I_s=-m_{43}$, $I_{c}=m_{33}$ of a clean Gasb surface
with approximately $7 \mbox{nm}$ of oxide. In addition it shows, as an
overview, the ellipsometric measurements of samples $A$, $B$ and $C$
(short cones) and $D$~(longer cones). All cones were formed by
sputtering at normal incidence. The nano-structuration of the surface
strongly modifies the polarisation dependent optical response. Another
interesting feature is the reflectance of such nanostructured
surfaces, which have additional practical applications. It is
particularly clear from figure~\ref{fig:R} that the reflectance is
much reduced, compared to the clean surface, at higher photon energies.
Furthermore, the antireflection properties tend to appear for lower energies as the cones get higher. This
could be explained as a "motheye" effect from the graded index of
refraction~\cite{motheye}.

From FEG-SEM and HR-TEM images, it is observed that the samples consist
of conical nanostructures of various sizes. For sufficiently small
cones, the surface can be treated as a thin film layer of effective
medium. This layer will be uniaxially anisotropic , since the cones
will show a different response to an electric field normal to the mean
surface, than to a field parallel to it. Anisotropic uniaxial
materials with the optic axis in the plane of incidence, will appear
like an isotropic material under ellipsometric investigations
$(r_{sp}=r_{ps}=0)$~\cite{Azzam}, in the sense that reflections from
such materials will be described by a diagonal Jones matrix, and by a
block diagonal Mueller-Jones matrix. This means that all the
polarisation altering properties of the structured surface can be
described by the ellipsometric angles $\Psi$ and $\Delta$, derived
from the ellipsometric intensities $I_s$ and
$I_c$.

Cones being directed normal to the surface have a
symmetry axis that is normal to the mean surface
, and the approximated effective media must therefore have
an optic axis in the same direction, {\it i.e.} it will appear like an
isotropic material, and can be fully characterised by
regular~(standard) ellipsometry. The samples will then have full
azimuth rotation symmetry (around the sample normal). If the cones are
tilted from the sample normal, this will generally no longer be the
case (except for the two special azimuth orientations where the tilted
cones lie in the plane of incidence). The structures will then
correspond to an anisotropic material with a tilted optic axis. To
describe the polarising properties of reflections from such a surface,
one also needs to account for the coupling of the $s$- and
$p$-polarisation through the reflections
coefficients $r_{sp}$ and $r_{ps}$, in addition to $r_{pp}$ and
$r_{ss}$. A long range ordering, or anisotropic shapes of the
individual cones, would also break the rotation symmetry and give
polarisation coupling. To fully characterise such a sample, one
needs to perform generalised ellipsometry (see
{\it e.g.}.~\cite{Jellison:97,Laskarakis}) or Mueller-matrix
ellipsometry. Mueller matrix ellipsometry has a great advantage over
generalised ellipsometry, since it also can deal with depolarizing
samples, which is not the case for the latter. Depolarisation may
arise from irregularities in the structure (shape, size and ordering)
and from multiple scattering. If the cones are small enough to be
treated by effective medium theory, the
structures will have the same effect as layers that are homogeneous in
a plane parallel to the surface, and there will be no multiple
scattering.  When the dimensions of the cones exceed the validity of
the effective medium theory, the inhomogeneities will give rise to multiple scattering and depolarisation. In this case there will
be coupling between the polarisation modes even though the structures
are rotationally symmetric and point normal to the surface, since the
structures no longer can be approximated as an effective homogeneous
layer. From this observation, one may conclude whether a given sample
can be modelled accurately by effective medium theory from
measurements of depolarisation alone.

% Slightly tilted cones will give rise to "anisotropic" deviations
% from zero in the "off-block-diagonal" elements of the Mueller
% matrix, and also break the azimuth rotation symmetry. We would also
% suspect that in the case where the effective medium theory is not
% valid, the Mueller matrix would become off diagonal even though the
% cones point normal to the surface. A long range ordering, or
% anisotropic shapes of the cones, could also give rise to
% non-block-diagonal Mueller matrices.

\begin{figure}
  \centering
  \includegraphics[width=0.5\textwidth]{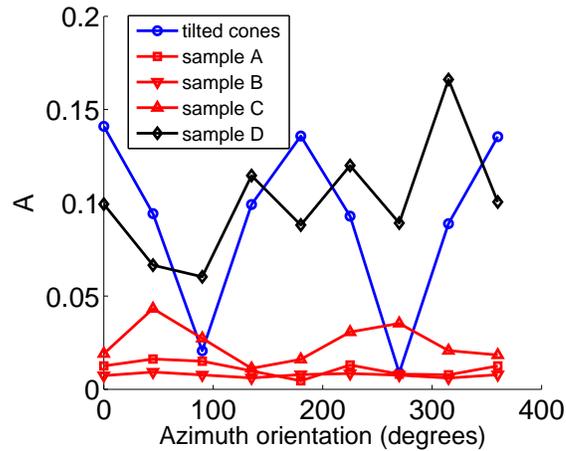}
  \caption{Degree of non-block-diagonality ($A$, as defined in equation
    (\ref{eq:DOA})) for various samples, as a function of azimuth
    sample rotation. The sample denoted tilted cones consisted of cones
    tilted $45^\circ$ from the surface normal (approx. $30 \mbox{nm}$
    high), while the other samples have cones pointing normal to the
    surface.}
  \label{fig:A}
\end{figure}

To examine if the samples give polarisation coupling, their Mueller
matrices measured by MME have been analysed. If there is no coupling,
the Mueller matrix should be block-diagonal.
%Indeed, the longer cones showed considerable "anisotropy" (non-block diagonal Muller matrices) for certain orientations, and also considerable depolarization. The anisotropy was inspected by Mueller Matrix Ellipsometry (MME) in the visible range.
We define a measure of the degree of
non-block-diagonality as
\begin{equation}
A=\left(\frac{m_{13}^2+m_{14}^2+m_{23}^2+m_{24}^2+m_{31}^2+m_{32}^2+m_{41}^2+m_{42}^2}{m_{11}^2+m_{12}^2+m_{21}^2+m_{22}^2+m_{33}^2+m_{34}^2+m_{43}^2+m_{44}^2}\right)^{1/2},
\label{eq:DOA}
\end{equation}
which is $0$ for block-diagonal Mueller matrices (such as reflections
from isotropic surfaces), and has the value $1$ for maximum non-block-diagonal matrices (such as circular polarisers
and $\pm45$ linear polarisers). Figure~\ref{fig:A} shows this quantity
as a function of azimuth sample rotation around the mean surface
normal for various samples. Additionally, as a reference, a sample
with nanostructures sputtered at $45$ degrees of incidence, with an
effective layer thickness of approximately $30 \mbox{nm}$ is also
shown~\cite{ICSE-4,Nerbotilt}. This sample consists of cones tilted by
approximately $45$ degrees from the mean surface normal, and has as
expected a Mueller matrix that is only block-diagonal
for azimuthal orientations where the cones lie in the plane of
incidence.

Moreover, it is observed from figure~\ref{fig:A}, that the short
cone samples have negligible polarisation coupling, while the longer
cones have substantial deviations from a block-diagonal Mueller
matrix. This coupling could, as earlier discussed, be related to a
slight tilting of the cones, to a long range preferential ordering of
the cones, or an anisotropic shape of the individual cones. It is
speculated that a long range preferential ordering could be induced by
{\it e.g.} substrate polishing features. However, for sample $D$, no azimuthal orientation has been observed to give a
block-diagonal Mueller matrix, as should be the case for a thin film
with the optic axis in the plane of incidence. This implies that these
samples cannot be modelled as an anisotropic thin film layer, and
that their optical properties are strongly affected by multiple
scattering. Such samples can not be fully characterised by SE, and full Mueller matrix ellipsometry is instead necessary.
The samples $A$, $B$ and $C$ only show a slight deviation
from block-diagonal Mueller matrices, and these off-diagonal elements will be neglected in the
following analysis and modelling. The detailed
analysis and modelling of tilted cones will be treated in a separate
publication~\cite{Nerbotilt}.
 
\begin{figure}
  \centering
  \includegraphics[width=0.5\textwidth]{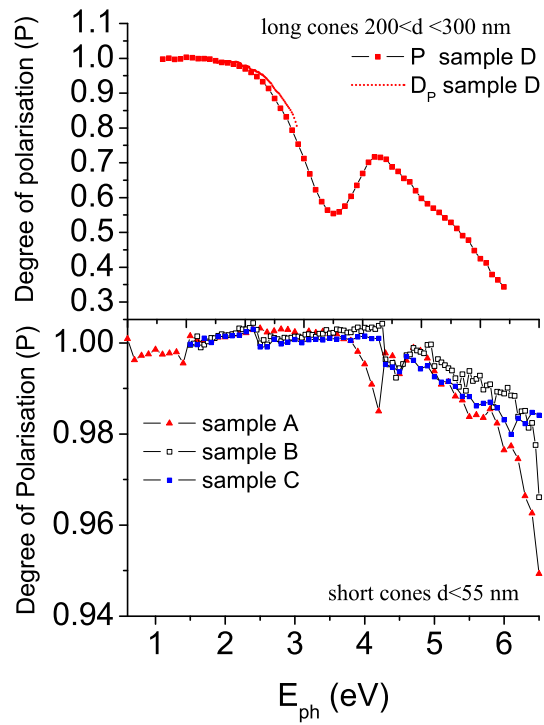}
  \caption{Degree of polarisation (P) as calculated from the PMSE
    measurements by equation (\ref{DOP}), for the nanostructured GaSb
    samples. The bottom figure shows P for the short cone samples: $A$,$B$
    and $C$, while the top figure shows P for the long cones of sample
    $D$. The depolarisation index $D_P$, calculated from the Mueller
    matrix in the visible range, is also shown for the long cones in
    the top figure.}
  \label{fig:DOP}
\end{figure}

From the polarisation coupling at various azimuth orientations
of sample $D$, it was concluded that the polarisation altering
properties of this sample had contributions from multiple scattering,
and that it would not be well approximated as an effective media. From
this conclusion, one would expect the sample to be depolarising, which
is confirmed by the depolarisation index ($D_P$, defined in
equation~(\ref{eq:DP})) evaluated from the MME measurements (figure
\ref{fig:DOP}). As expected, the depolarisation increase for
increasing photon energy, since the effective medium approximation
gets less accurate for decreasing wavelength. In addition, an
approximation to the depolarisation at higher energies has been found
by calculating the degree of polarisation, $P$, from the PMSE
measurements through equation~(\ref{DOP}). The degree of polarisation
obtained in this way is a measure of how much certain polarisation
states are depolarised, and will generally differ from the
depolarisation index, which (in many cases) is the average
depolarisation of all possible incident polarisation states
(see~\cite{Chipman:05}). For samples with block-diagonal Mueller
matrices ($A$, $B$, $C$), the degree of polarisation can safely be
used as a measure of depolarisation. It is observed that the short
cones have principally a low depolarisation throughout the measured
spectral range (figure~\ref{fig:DOP}). All the samples studied in the present work show an increasing depolarisation towards the UV
range. Sample $A$ has a small dip in the degree of polarisation at the
photon energy where $I_s=0$ and $I_c=1$. This effect can be explained
by a small variation in cone height (thin film
thickness)~\cite{Jellison-McCamy} or cone shape. It could also be
caused by quasi-monochromatic light from the monochromator.  It is
especially noted that the dielectric function is descending steeply at
this photon energy~\cite{Aspnes}, meaning that a very small wavelength
distribution could give depolarisation. It is noted that samples $B$
and $C$, show little depolarisation in the main part of the spectrum.
This does not imply that these samples have less variation in cone
height or shape than sample $A$, since there is no photon energy for
which $I_s=0$ and $I_c=1$~(see figure \ref{fig:IsIc}). All the short
cones still show a small but observable increasing depolarisation for
decreasing wavelength. Furthermore, it is observed that the degree of
polarisation, $P$, decreases more rapidly as a function of
wavelength as the cone height increases.

\subsection{Optical modelling}

The cones with no or little depolarisation ("short-cones") have been
modelled as a graded anisotropic thin film layer of effective media,
on a GaSb substrate. Reflection coefficients have been calculated by
an implementation of Schubert's algorithm~\cite{Schubert} for
reflections from arbitrarily anisotropic layered systems, based on
Berreman's $4\times 4$ differential matrices~\cite{Berreman:72}. As a
first approximation, the cones have been modelled as a stack of
cylinders with decreasing diameter. Each cylinder in the stack defines
a layer with a homogeneous effective dielectric function. With a
sufficient number of layers, this will be a good approximation of a
graded thin film layer. Based on HR-TEM observations, we have assumed
the cylinders to consist of a core of crystalline GaSb, covered by a
coating consisting of a mixture of amorphous GaSb and GaSb oxide. The
anisotropy is introduced by using the generalised Bruggeman effective
medium theory~\cite{Spanier}, giving the formula

\begin{eqnarray}
  f_{c-GaSb}\frac{\epsilon_{c-GaSb}-\epsilon_{ii}}{\epsilon_{ii}+L_i(\epsilon_{c-GaSb}-\epsilon_{ii})}&+&f_{coat}\frac{\epsilon_{coat}-\epsilon_{ii}}{\epsilon_{ii}+L_{i}(\epsilon_{coat}-\epsilon_{ii})}\nonumber
  \\
  &~&+f_{void}\frac{\epsilon_{void}-\epsilon_{ii}}{\epsilon_{ii}+L_{i}(\epsilon_{void}-\epsilon_{ii})}=0,
\label{G-BR}
\end{eqnarray}
where $f$ and $\epsilon$ denote the filling factors and complex
dielectric functions, respectively, with the subscript $c-GaSb$
referring to the crystalline core, $coat$ to the coating over layer,
and $void$ to the surrounding void. $L_{i}$ denotes the depolarisation
factor in direction $i$ (along a principal axis of the structure) and $\epsilon_{ii}$ is the effective
dielectric function in direction $i$. Our principal axes will be two orthogonal axes parallel to the mean surface, $x$ and $y$, and a $z$ axis normal to the mean surface. The dielectric function of the
coating, $\epsilon_{coat}$, has been determined by letting it be 
a mixture of amorphous GaSb~(a-GaSb) and GaSb oxide~(oxide), and 
using the standard Bruggeman equation for spherical inclusions ($L_i=1/3$)
\begin{equation}
  f_{a-GaSb}\frac{\epsilon_{a-GaSb}-\epsilon_{coat}}{\epsilon_{coat}+2\epsilon_{a-GaSb}}+f_{oxide}\frac{\epsilon_{oxide}-\epsilon_{coat}}{\epsilon_{coat}+2\epsilon_{oxide}}=0
\label{BR}
\end{equation}
These cylinders can thus be approximated as an effective thin film
layer, which is valid when the distance between neighbouring cylinders
are much smaller than the wavelength of the light. The layer will be
anisotropic, with depolarisation factor $L_x=L_y=0.5$ in the plane
parallel to the surface, and $L_z=0$ in the direction normal to the
surface. The reflection coefficients from such an anisotropic layered
system has been calculated, and used to find the ellipsometric
intensities $I_s$ and $I_c$~\cite{Hauge}
\begin{equation}
  I_s=\frac{2 Im(r_{pp}r_{ss}^*+r_{ps}r_{sp}^*)}{|r_{ss}|^2+|r_{pp}|^2+|r_{sp}|^2+|r_{ps}|^2}
\end{equation}	

\begin{equation}
  I_c=\frac{2 Re(r_{pp}r_{ss}^*+r_{ps}r_{sp}^*)}{|r_{ss}|^2+|r_{pp}|^2+|r_{sp}|^2+|r_{ps}|^2}
\end{equation}

The parameters of the models have been fitted to $I_s$ and $I_c$ by
minimising $\chi^2$, defined as
\begin{equation}
\chi^2=\frac{1}{2N-M+1}\sum_{i=1}^N\left( \frac{(I_{si}^{mod}-I_{si}^{exp})^2}{\sigma_{si}^2}+\frac{(I_{ci}^{mod}-I_{ci}^{exp})^2}{\sigma_{ci}^2}\right)
\label{chi}
\end{equation}
where N and M are the number of measurement points the number of free
parameters in the model, respectively. $\sigma_{si}$ and $\sigma_{ci}$
are the standard deviations of the respective measurements. Additional
measurements such as any Mueller matrix element, or the reflectance,
may be added to the formulae in a similar fashion.

\begin{figure}
  \centering
  \includegraphics[width=0.5\textwidth]{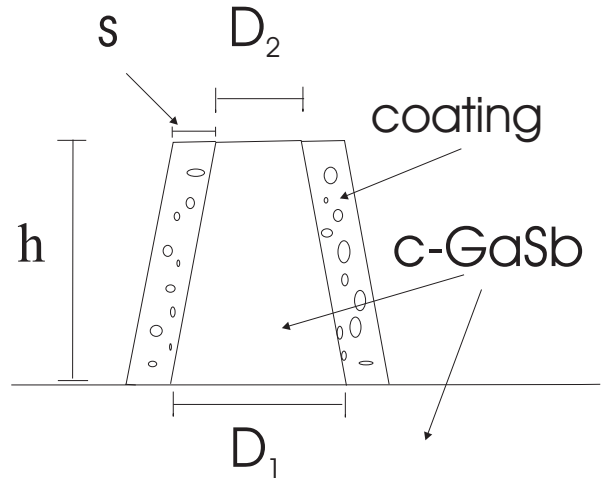}
  \caption{Parameters used in the graded effective medium model. $h$ is the total height, $s$ is the thickness of the coating of amorphous material and oxide, $D_1$ and $D_2$ are lower and upper diameters of the crystalline core.}
  \label{fig:param}
\end{figure}

The simplest model giving satisfying results has been one with 5
parameters (see figure~\ref{fig:param}), the total height $h$, the
relative (effective) diameters $D_1$ and $D_2$ of the bottom and top
cylinder cores, the thickness of the coating $s$, and the amount of
oxide in the coating, $f_{oxide}$. The two diameters $D_1$ and $D_2$,
and the thickness $s$, are dimensionless quantities, defined as
fractions of the mean nearest neighbour distance of the cones. This
distance can not be found from the optical measurements when the
effective medium approximation is valid, since the effective medium
only depends on volume filling factors and shape through the
depolarisation factors $L_i$. This means that the model is independent of the scale in the horizontal plane, for all structures sufficiently
smaller the the wavelength of light. A stack of $\mathcal{N}$=100
cylinders of equal height were used to approximate a continuous
gradient, with the diameters $d(n)$ of layer $n$ decreasing
linearly from $D_1$ to $D_2$:
\begin{equation}
  d(n)=D_1-\frac{D_1-D_2}{\mathcal{N}-1} n
\end{equation}	
for $n=0,1,\ldots,99$. Assuming a hexagonal ordering of the cylinders, the filling factor of crystalline GaSb and coating become:
\begin{equation}
%  f_{c-GaSb}(n)=\frac{2\pi d^2(n)}{\sqrt{3}\, 4},
  f_{c-GaSb}(n)=\frac{\pi}{\sqrt{12}} \, d^2(n)
  \label{eq:filling}
\end{equation}
and
\begin{equation}
  f_{coat}(n)=\frac{2\pi}{\sqrt 3}(d(n)s+s^2).	
\end{equation}
Notice that for an effective medium theory it is only the filling factors that play a role, and not the specified ordering of the cones. As long as the filling factors remain the same, effective
medium theory can not distinguish between different geometrical arrangements.The distance between the centers of neighbouring cones has been set to unit length. The thickness of
coating is constant for all layers.

Minimisation was performed using the sequential quadratic programming
(SQP) algorithm of the Matlab Optimisation Toolbox 3.1.1. The
dielectric functions of crystalline GaSb, amorphous GaSb and GaSb
oxide, were obtained from the literature~\cite{Aspnes,aGaSb,Zollner}.
The standard deviations (noise) $\sigma_{si}$ and $\sigma_{ci}$ of the
ellipsometric measurements $I_s$ and $I_c$ were estimated to be 0.01.

The resulting parameters of the fitted models are given in
table~\ref{tab:model}. The model gave a good fit to the optical
measurements of sample $A$, with a cone height of $54 \mbox{nm}$, and
a clear grading in the inner cylinder diameters from $D_1=0.55$ to
$D_2=0.36$.  This is in good agreement with the previously presented
SEM and TEM images (see figure~\ref{fig:TEM_IM}).

Sample $C$ could be well fitted by a model with $D_1\approx D_2$,
meaning that it could have been modelled equally well by only one layer
of coated cylinders. The cone height was found to be $36
\mbox{nm}$. It may be that the optical measurements are not sensitive
to a possible gradient in such a short structure, or that the
structures have a shape resembling a cylinder.
 
The ellipsometric measurements of sample $B$ greatly resemble those of sample $C$,
but the optical model could not give an equally good fit. When $D_1$
and $D_2$ are allowed to vary freely, the model converges to a
seemingly unphysical case (based on the TEM and SEM images) with
$D_2>D_1$. To avoid this problem, they have been constrained so that
$D_1>D_2$. The result is then a cylinder like model (no grading), with
a height of $32 \mbox{nm}$. It may appear to be necessary to develop
more advanced models to perfectly fit the measurements of this sample.
Natural extensions could be to let the coating thickness $s$ vary with
height; letting the diameter \emph{d(n)} of layer $n$
follow a non-linear function from $D_1$ to $D_2$; or letting the filling factors be able to deviate from values consistent with a hexagonal ordering. We will, however, not treat such advanced models
here, but keep the parameters in the models to a minimum for easier
interpretation, and to avoid unphysical solutions. It is also
plausible that the dielectric functions of the different phases mixed
in the effective medium theory are somehow different, {\it e.g.} that
the properties of the oxide is different.

According to the results from the optical characterisation, sample $B$ should consist of slightly shorter cones than sample $C$. This seems overall consistent with the AFM observations. It has been observed for sample $A$, $C$ and $D$ that the nearest neighbour distance increase for increasing cone height (table \ref{tab:AFM}). From the cone $density$ one should therefore expect sample $B$ to have shorter cones than sample $C$. The average cone height $\langle h \rangle$ estimated from AFM did not show as clear a difference between the samples, but such small height differences could possibly be masked in the uncertainty in the estimation of $\langle h \rangle$.
 
%\del{Differences between sample $B$ and $C$ can also be seen from the AFM
%results presented in table~\ref{tab:AFM}. Sample $B$ was found to have
%a cone density of $948$ cones per $\mu m^2$, while $C$ had $766$ cones
%per $\mu m^2$. In accordance to the AFM observations, $B$ has a higher
%filling factor (total effective diameter of $0.72$) in the optical
%model than $C$ (total effective diameter of $0.63$), and the ratio
%between the densities and the ratio between the filling factors is
%approximately the same.}

The height of the cones of sample $B$ and $C$ obtained from the
optical model, are lower than the average heights found by AFM. It
should be stressed that the height of sample $A$ (which coincided well
with the height from the optical model) were found in a different way
(by HR-TEM).
%It has been reported that AFM images tend to exaggerate
%the height of the patterns~\cite{ChanAFM},
As previously mentioned, the average cone height estimated from AFM measurements may be exaggerated.
% which may be one explanation for the height discrepancy from the optical model. 
The model appears to be very sensitive to changes in the thickness of the
effective medium layer, a perturbation in thickness of only a few $\mbox{nm}$
results in a large increase of $\chi^2$. However, different models may
result in different layer thicknesses. For instance, it may seem more
reasonable to let the cones be covered by a coating of thickness $s$
also on the top. This has been tested, and resulted in equally good
$\chi^2$ values as the models reported in table~\ref{tab:model}, but
with total heights $4$--$5 \mbox{nm}$ higher. The problem with such a
model, is that the thickness of the coating top layer has to be
determined absolutely, not just as a ratio, $s$, of the nearest
neighbour distance. This distance can not be obtained from SE
measurements, but must be found from {\it e.g.} AFM or SEM studies. We are
interested in a model that can help us characterise the nanostructures
from SE measurements alone, and therefore reject this model with a
coating also on the top.

The total volume filling factors for the optical models are tabulated in the last column of table \ref{tab:model}. For ideal cones, ordered in a hexagonal lattice, the maximal filling factor is $0.30$. The model filling factors lie in the range $0.34$--$0.46$, in good correspondence to the rounded conical structures observed from TEM, SEM and AFM measurements (rounded cones will give a larger filling factor than cones with a sharp top). 
Exact estimation of filling factors from microscopy images proved to be difficult. The varying shape and size of the individual cones must be taken into account, together with the mean nearest neighbour distance. The AFM measurements should in principle be ideal for this, but because of a too blunt tip and "`holes"' in the surface, they drastically overestimate the filling factors. By estimating the shape and size of an individual cone from a TEM image, and using the mean nearest neighbour distance from AFM measurements, a rough estimation of the filling factor of sample A was found to be $0.36\pm0.04$, in reasonable agreement with the value from the optical model ($0.39$).
% Problem: the thickness of the coating top layer in such a model has
% to be determined absolutly, not just a as ratio s of then nearest
% neibour distance. This can not be done from SE measurements, but
% must be found from AFM or SEM studies. (We used the <D> values from
% AFM tabulated in tabel.. . ) To get a measuemremet method based on
% SE only, we do not chose such a model... inconsistent... bla bla. Be
% aware that not including cotaing overlayer might underestimate the
% heights (around
% 10%, depending on coating thickness), this underestimation will 
% probalby be quite systematic, i.e samples fitted to the same model
%v will return heights realtivly correct to each other...
 
\begin{figure}

  \begin{minipage}[b]{0.5\linewidth} % A minipage that covers half the page
    \centering
    \includegraphics[width=\linewidth]{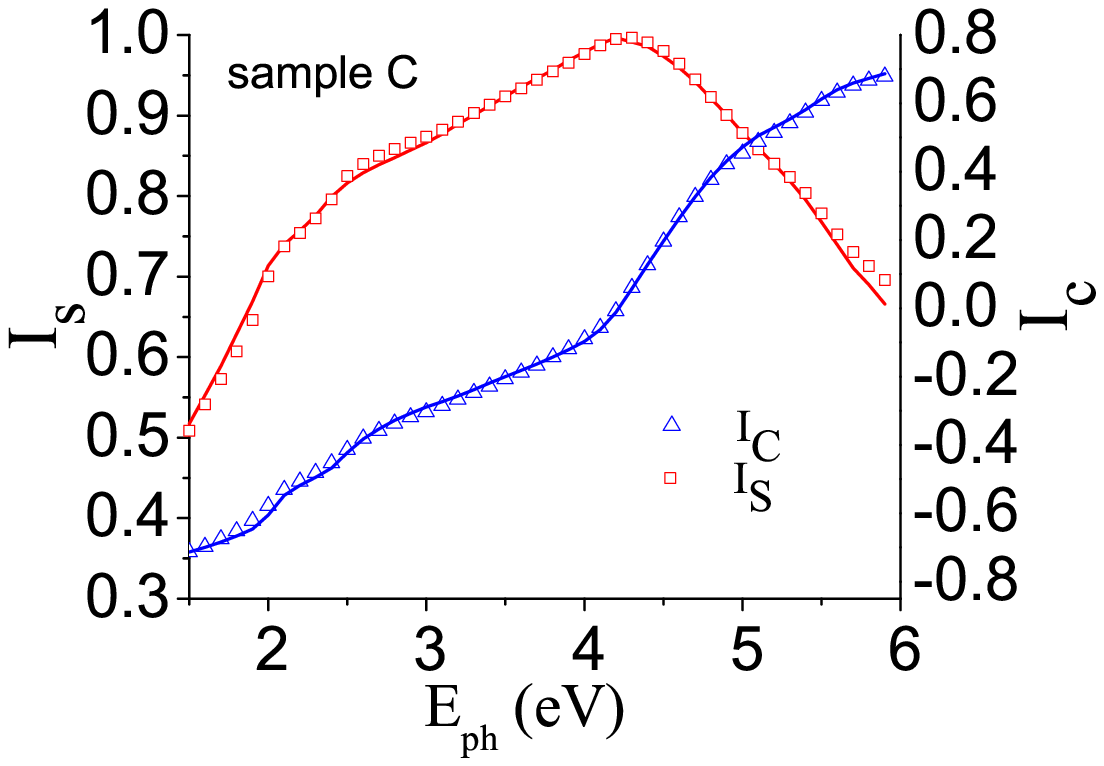}
  \end{minipage}
  % \hspace{0.5cm} % To get a little bit of space between the figures
  \begin{minipage}[b]{0.5\linewidth}
    \centering
   \includegraphics[width=\linewidth]{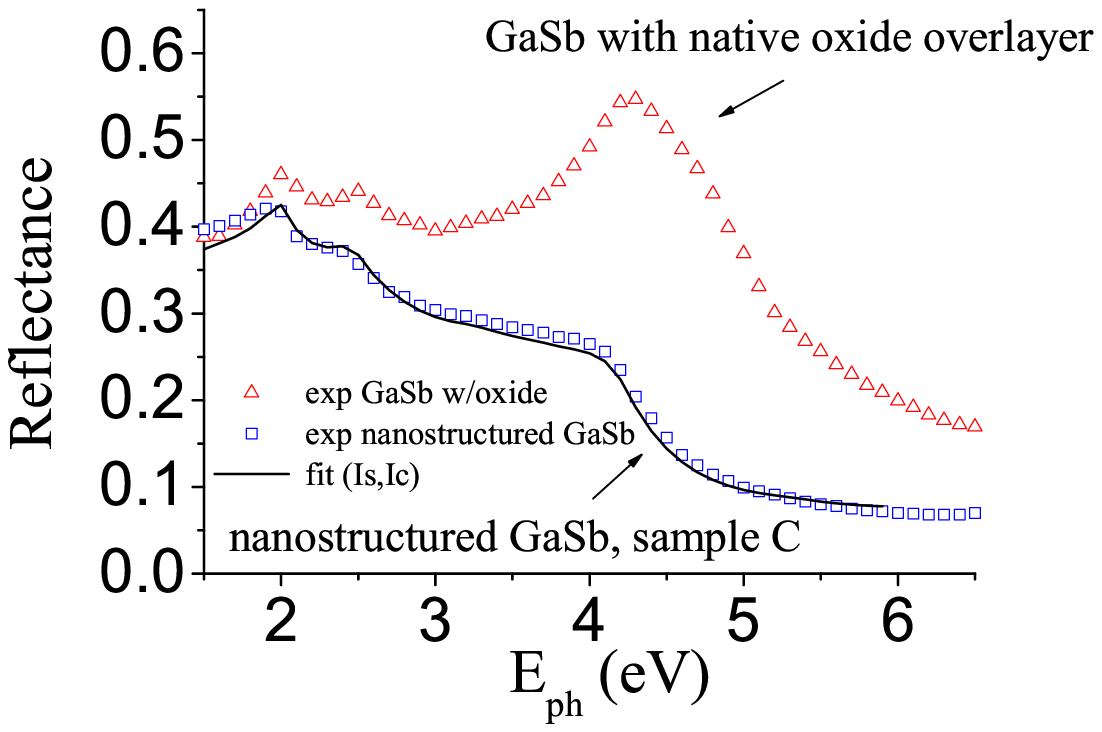}
 \end{minipage}
 \caption{Fitted model to the ellipsometric measurements
   $I_s=-m_{43}$, and $I_c=m_{33}$, for sample $C$(left). The right
   figure depicts the measured reflectance $R=(|rss|^2+|rpp|^2)/2$, and
   simulated reflectance, calculated from the fitted model
   parameters.}
 \label{fig:n113}
\end{figure}

The construction of an effective medium optical model, predicts structural parameters that correspond reasonably well to the physical height of the samples, and the density/shape of the
cones. Equally important, the models can by used to predict optical
properties not measured. The model of sample $A$, calibrated by SE
measurements at $55^\circ$ angle of incidence, were used to
successfully predict results of SE measurements at $70^\circ$ angle of
incidence. The models may also predict the reflectance ($R_{ss},
R_{pp}$ or $R$) of the samples (see figure~\ref{fig:n113}). We propose
that the models can be used as a tool to calculate the polarisation
dependent optical properties of such samples at any angle of
incidence.

% The reflectance of sample $C$ (at $55^\circ$ angle of incidence)
% were calculated from the model, and were found to

% This model calibrated to SE measurements at 55$^\circ$ angle of
% incidence, were used to successfully predict results of SE
% measurements at 70$^\circ$ angle of incidence. We propose that it
% can be used as a tool to calculate the polarization dependent
% optical properties of such samples at any angle of incidence.
	
% However, it is equally important that the calibrated optical model
% may reproduce other parameters, such as reflectance ($R_{ss},
% R_{pp}$ or $R$) at an arbitrary angle of incidence. The models
% fitted to $I_s$ and $I_c$ were tested by calculating the
% reflectance, and comparing it to an experimentally determined
% reflectance, see figure \ref{fig:R}. The figure also shows as a
% reference, the measured spectrum from plane GaSb covered with a 5-7
% nm oxide over layer.

The large depolarising properties of sample $D$ indicate that it may
not be modelled appropriately by effective medium theory over the full
spectral range considered. The experimental observations represent an
interesting case, in the sense that there are no commonly available
models to appropriately fit these data. A tentative effective medium
modelling between $0.6$ and $2.5 \mbox{eV}$ was tested in order to
extract approximately the cone height of sample $D$. It was found to
be 165 nm, about half of the height found by SEM, but still
considerably higher than heights found for the short cones, and with a
clear gradient. The dielectric function data for GaSb-oxide and c-GaSb
in the photon energy range $0.6$--$1.5 \mbox{eV}$, were not available
in the literature, and were therefore extrapolated from PMSE
measurements at $70^\circ$ angle of incidence. The parameters of the
resulting model are tabulated in order for completeness. Improved
optical models suitable for modelling of the optical response of these
samples are currently being undertaken, and planned for future work.

\section{Summary and conclusions}

Spectroscopic ellipsometry and Mueller matrix ellipsometry have been
shown to be useful techniques for the characterisation of
nanostructured surfaces, such as nanocones of GaSb on GaSb. Overall, the
observations from SE appear to be consistent with the results
from SEM, TEM and AFM studies. An optical model has been found to fit
well to the measurements obtained for short cones (of height $55
\mbox{nm}$ and lower).This was achieved by treating the structures as
a graded anisotropic thin film of effective medium. These models have
been applied in order to obtain an approximation to the average cone
height of the samples, and also, to some extent to gain information on
the cone shape. They may also be used to estimate reflectance and
polarisation altering properties for reflections at any angle of
incidence. The nanostructuration of the surface was shown to
considerably reduce the reflectance. The antireflecting properties
increased with cone height. Samples with long nanocones ($200$--$300
\mbox{nm}$) were found to be strongly depolarising, and could not be
modelled as an effective medium. The full Mueller matrix must be
measured to fully characterise the polarisation altering properties of
such samples.  We have demonstrated that SE can be a fast and
non-destructive way of characterising nanocones of GaSb, with the
possibilities of \emph{in-situ} control under production.

\ack

The authors are grateful to M.~Stchakovsky at Horiba Jobin Yvon
for access to scientific instruments, and Susanne W.~Hagen at NTNU for
doing complementary measurements.

% --------------------------------------------------------------------
% BIBLIOGRAPHY
% --------------------------------------------------------------------
\section*{References}

\bibliographystyle{iopart-num} % (uses file "plain.bst")
\bibliography{elliref_v4h} % expects file "myrefs.bib"

\begin{thebibliography}{10}
\bibitem{Nerbotilt}
{Nerb\o} I~S, {Kildemo} M, {Hagen} S~W, {Leroy} S and {S\o nderg\aa rd} E
  Optical properties and characterisation of tilted gasb nanocones (To be
  published)
\end{thebibliography}

\clearpage

\begin{table}
\caption{\label{tab:samples}Sputtering conditions and definition of the samples. All samples are sputtered at normal incidence, with no temperature control.} 

\begin{indented}
\lineup
\item[]\begin{tabular}{@{}*{5}{c}}
\br                              
Sample name &Sputter Time&Mean Temperature&Applied Voltage&Average Flux\cr
 & (min)&($^\circ$C)&(V)&$(mA/cm^2)$ \cr
\mr
$A$&	$10$&	$33$&	$-400$&	$0.098$\\
$B$	&$10$&	$41$&	$-400$&	$0.096$\\
$C$&	$10$&	$35$	&$-300$&	$0.28$\\
$D$&	$10$	&$47$&	$-500$&	$0.37$\\
\br
\end{tabular}
\end{indented}
\end{table}

\begin{table}
\caption{\label{tab:AFM}Results of AFM,TEM and SEM studies of GaSb nanocone samples. $\left<h\right>$ is the average cone height, with standard deviation $\sigma_h$. $Density$ is the number of cones pr. $\mu m$, $\left<D\right>$ is the average distance between neighbouring cone centers, while $\alpha$ is the average cone angle. The tabulated results have been found from; $^1$ TEM studies, $^2$ AFM measurements and $^3$ SEM images.} 

\begin{indented}
\lineup
\item[]\begin{tabular}{@{}*{6}{c}}
\br                              
Sample name&	$\left<h\right>$ &	 $\sigma_h$ &	Density&	$\left<D\right>^*$&$\alpha$ \\
 & (nm)&(nm)&($\mu m^2$)&(nm)&degrees \cr
\mr
$A$&	$55^1$&	$5.4^1$&	$549^2$&	$46^2$	&$73^1$\\
$B$	&$46.5^2$&	$5.2^2$&	$948^2$&	$35^2$&	\\
$C$&	$47.6^2$&	$8.85^2$&	$766^2$	&$39^2$&	\\
$D$&$299^3$	&$40^3$&	$74.25^2$&	$125^2$&	$77.2^3$\\
\br
\end{tabular}
\end{indented}
\end{table}

\begin{table}
  \caption{\label{tab:model}Parameter results from the fitting of the optical models to the ellipsometric data. $h$ is the total height of the model layers, $D_1$ and $D_2$ is the bottom and top diameter of the crystalline core, $s$ is the coating thickness, and $f_{oxide}$ is the amount of oxide in the coating. $\chi^2$ is the square deviation of the modelled ellispometric intensities from the measured, as defined in equation (\ref{chi}). $D_1$,$D_2$ and $s$ are defined relative to the center to center distance for two nearest neighbours. The sample denoted by "*", was only curve fitted below $2.5\mbox{eV}$ ({\it i.e.} for $P>0.9$).} 

  \begin{indented}
    \lineup
  \item[]\begin{tabular}{@{}*{8}{c}} \br
      Sample name&	$h$ &	$D_1$&	$D_2$&	$s$&	$f_{oxide}$&	$\chi^2$&$f_{tot}$\\
      & (nm)&&&&& \cr \mr
      $A$	&$54$&	$0.55$& $0.36$&	$0.10$&	$0.56$&	$2.6$&$0.39$\\
      $B$	&$32$&	$0.31$&	$0.31$&	$0.21$&	$0.64$&	$12.4$&$0.46$\\
      $C$&	$36$&	$0.36$&	$0.35$&	$0.14$&	$0.34$&	$1.1$&$0.37$\\
      $D^*$	&$165$&	$0.95$&	$0.21$&	$0$	&$0.0$	&$7.4$&$0.34$\\
      \br
    \end{tabular}
  \end{indented}
\end{table}

%\begin{figure}
%  \centering
%  \includegraphics[width=0.5\textwidth]{TEM1microtomySmall}
%  \caption{HR-TEM image of sample $A$, prepared by ultramicrotomy.}
%  \label{fig:TEM_M}
%\end{figure}

\begin{figure}

  \begin{minipage}[b]{0.5\linewidth} % A minipage that covers half the page
    \centering
    \includegraphics[width=\linewidth]{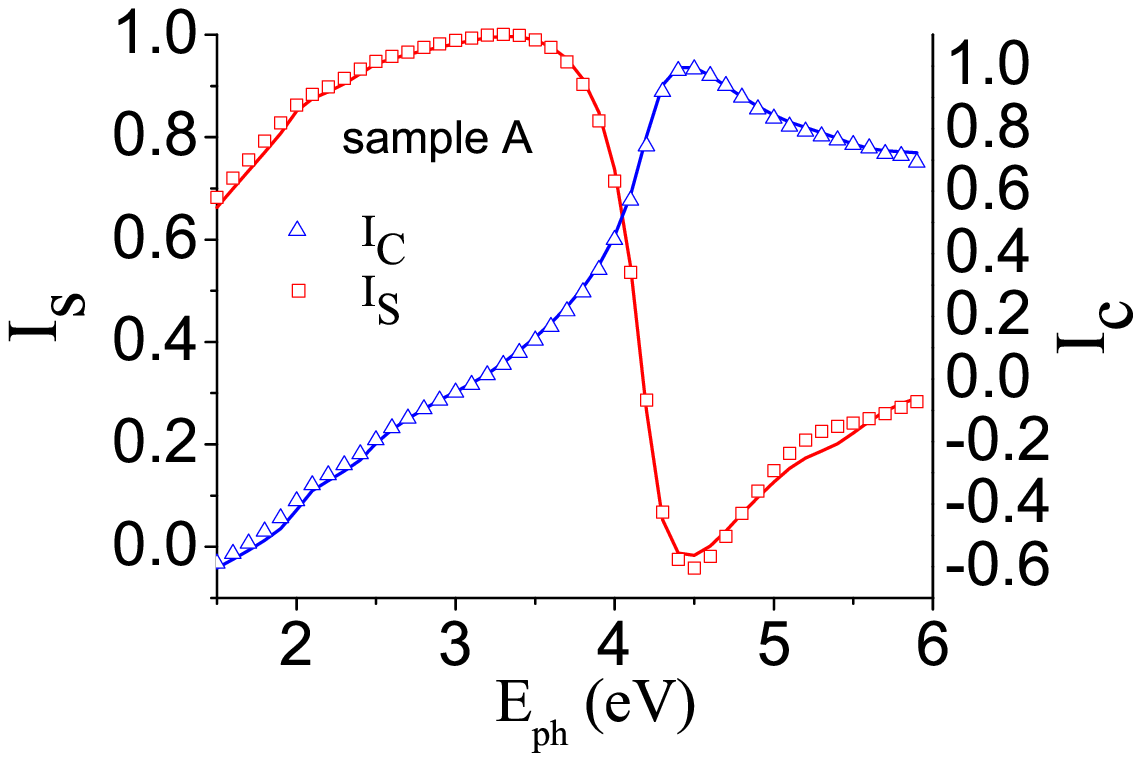}

  \end{minipage}
  % \hspace{0.5cm} % To get a little bit of space between the figures
  \begin{minipage}[b]{0.5\linewidth}
    \centering
    \includegraphics[width=\linewidth]{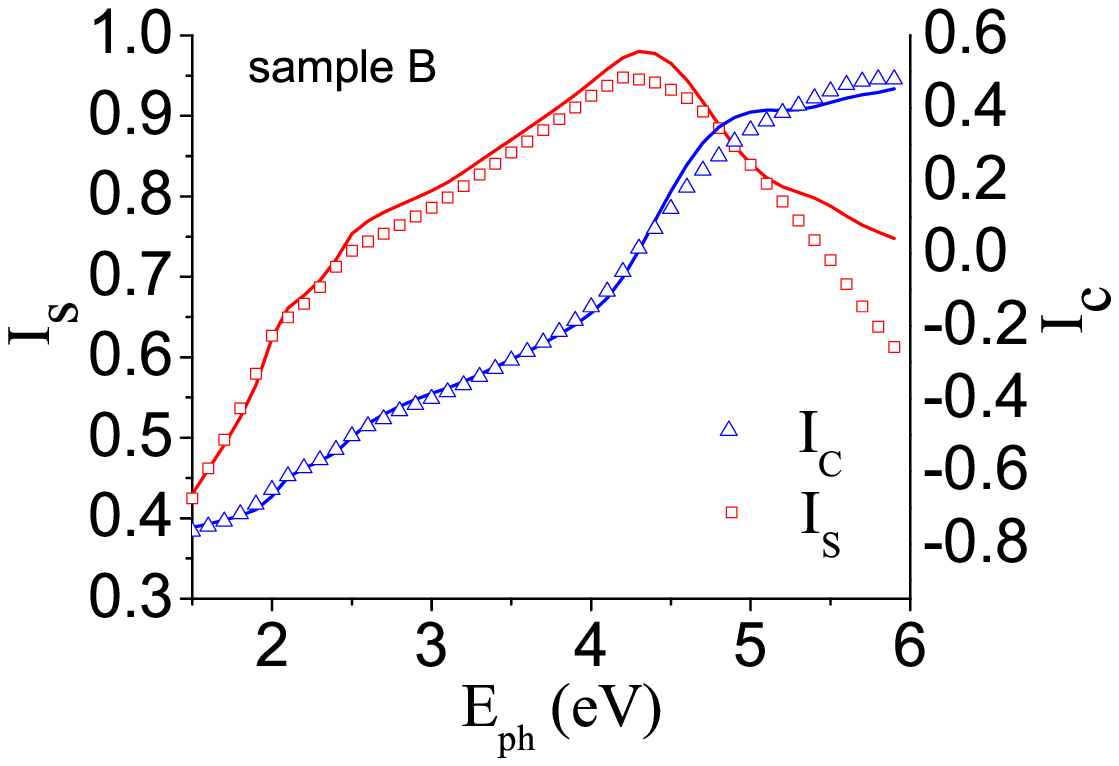}

  \end{minipage}
  \caption{Fitted model to the ellipsometric measurements
    $I_s=-m_{43}$, and $I_c=m_{33}$, for sample GaSb $A$ (left) and
    sample $B$ (right)}
\end{figure}

\end{document}